\documentclass[12pt]{article}
\usepackage{amsfonts}
\usepackage{amssymb}
\usepackage{graphicx}
\newcounter{rown}
\def\bl{\setcounter{rown}{\value{equation}}
\stepcounter{rown}\setcounter{equation}0\def\theequation{\arabic{rown}\alph{equation}}}
\def\el{\setcounter{equation}{\value{rown}}
        \def\theequation{\arabic{equation}}}
\def\siz{\small}

\begin{document}
\title{ACCELERATION-ENLARGED SYMMETRIES IN NONRELATIVISTIC SPACE-TIME WITH
A COSMOLOGICAL CONSTANT\footnote{Supported by the KBN grant 1P03B01828}}
\author{J. Lukierski$^{1)}$, P.C. Stichel$^{2)}$  and W.J.
Zakrzewski$^{3)}$
\\
\siz $^{1)}$Institute for Theoretical Physics,  University of
Wroc{\l}aw, \\ \siz pl. Maxa Borna 9,
 50--205 Wroc{\l}aw, Poland\\
\siz and \\
\siz   Department of Theoretical Physics, Univ. of Valencia,\\
 \siz  46100 Burjassot (Valencia), Spain\\
 \siz e-mail: lukier@ift.uni.wroc.pl\\
\\ \siz
$^{2)}$An der Krebskuhle 21, D-33619 Bielefeld, Germany \\ \siz
e-mail:peter@physik.uni-bielefeld.de
\\ \\ \siz
$^{3)}$Department of Mathematical Sciences, University of Durham, \\
\siz Durham DH1 3LE, UK \\ \siz
 e-mail: W.J.Zakrzewski@durham.ac.uk
 }

\date{}
\maketitle

\begin{abstract}

By considering the nonrelativistic limit of de-Sitter geometry one obtains the nonrelativistic
space-time with a cosmological constant and Newton-Hooke (NH) symmetries. We show that the NH symmetry algebra can be 
enlarged by the addition of the constant acceleration generators and endowed with central extensions (one in any 
dimension (D) and three in D=(2+1)). 
We present a classical Lagrangian and Hamiltonian 
framework for constructing models quasi-invariant under enlarged NH
symmetries which depend on three parameters described by three nonvanishing central charges.
The Hamiltonian dynamics then splits into external and internal sectors with new non-commutative structures of external and internal 
phase spaces.
We show that in the limit of  vanishing cosmological constant the system reduces to the one presented in [1]
which possesses accelaration-enlarged Galilean symmetries.

\end{abstract}

\section{Introduction}

The two nonrelativistic Newton-Hooke (NH) cosmological groups were introduced by Bacry and Leblond 
in [2] who classified all kinematical groups in $D=(3+1)$. The NH symmetries can be obtained, in the limit $c\rightarrow \infty$, from the de-Sitter and anti-de-Sitter geometries [2-6]  and they describe, respectively, the nonrelativistic
expanding (with symmetry described by the $NH_+$ algebra) and oscillating (with symmetry described by the $NH_-$ algebra) universes. The cosmological constant describes the time scale $\tau$ determining the rate of expansion or the period of oscillation of the universe. When $\tau\rightarrow\infty$ we obtain the Galilean 
group and the standard flat nonrelativistic space-time. It has been argued that the NH symmetries and 
 the corresponding NH space-times can find an application in nonrelativistic cosmology [7,8,9] or even in M/String theory [10].

Recently,  we have considered in [1] the acceleration-enlarged Galilean symmetries\footnote{In [1] we described
our procedure of adding constant acceleration generators as an `extension'. To avoid confusion with the 
mathematical terminology  ($G'$ is an extension of $G$ by $K$ if $G=\frac{G'}{K}$) we call, in this paper, our procedure  an
`enlargement'.} with their central extensions and their 
dynamical realisations. The aim of this note is to show that these results, corresponding to $\tau\rightarrow\infty$, can be
generalised to the NH spaces with a finite cosmological constant ($\tau$ finite). In particular we show that the acceleration-enlarged NH symmetries,
as in the Galilean case, can have
\begin{enumerate}
\item for arbitrary $D$ - one central extension,
\item for $D=(2+1)$ - three central extensions.
\end{enumerate}
We also generalise to $\tau<\infty$ the actions providing the dynamical realisations of the NH algebra and the 
corresponding equations of motion.

The plan of our presentation is as follows. First, in Section 2 we recall some known results [5,6,8,11] on the 
NH algebras with central extensions.  We also add the constant acceleration generators, consider the most
general central extensions and describe the Casimirs of our new algebras. In Section 3 we present a (2+1) dimensional classical 
mechanics Lagrangian model with higher derivatives providing a dynamical realisation of the introduced symmetries.
In section 4 we consider its phase space formulation - with five 2-vector coordinates $(x_i,p_i,v_i,q_i,u_i)$.
In the following section we decompose its dynamics into two sectors (``external'' and ``internal''), with the six-dimensional
external sector describing a new $D=2$ non-commutative extended phase space. The last section contains some final remarks.

% Section 4 
%contains our conclusions and some remarks on the possibility of avoiding the ghost problem in models described by the field
%equations with higher derivatives. 

\section{ Acceleration-enlarged Newton-Hooke symmetries %($\widehat {NH}_+$ and $\widehat{NH}_-$)
 with central charges - algebraic 
considerations}

Let us first recall the ${NH}_+$ algebra - corresponding to an expanding universe.
In this case the Galilean algebra with the following nonvanishing commutators $(i,j,k,l=1,..D-1)$:
\bl
\begin{eqnarray} \label{onea}
[J_{ij},\, J_{kl}]\,&=&\,\delta_{ik}J_{jl}\,-\,\delta_{jl}J_{ik}\,+\,\delta_{jk}J_{il}\,-\,\delta_{il}J_{jk},\\
\label{oneb}
[J_{ij},\, A_{k}]\,&=&\,\delta_{ik}A_{j}\,-\,\delta_{jk}A_{i}\quad (A_i=P_i,K_i),\\
\label{onec}
[H,\,K_i]\,&=&\,P_i
\end{eqnarray}
is supplemented by the additional commutators\footnote{We pass to the $NH_-$ algebra by the substitution 
$\frac{1}{R^2}\rightarrow -\frac{1}{R^2}$ in (\ref{oned}).}
\begin{equation}\label{oned}
[H,\,P_i]\,=\,{K_i\over R^2}
\end{equation} \el
obtained by the deformation of the Galilean relations $[H,\,P_i]=0$ [12].

In arbitrary dimension $D$, as in the Galilean case, one can introduce a central extension describing the mass $m$\footnote{In this paper we represent our central generators as $\lambda{\bf 1}$ where $\lambda$ is a number and omit the unity operator ${\bf 1}$.}
\begin{equation}
\label{two}
[P_i,\,K_j]\,=\, m\delta_{ij}.
\end{equation}
In $D=(2+1)$ one can have a second central charge $\theta$ [6,8,11]
\bl
\begin{equation}
[K_i,\,K_j]\,=\,\theta \epsilon_{ij}.\label{threea}
\end{equation}
Algebraic consistency then requires that
\begin{equation}
[P_i,\,P_j]\,=\,-\frac{\theta}{R^2} \epsilon_{ij}.\label{threeb}
\end{equation}
\el

The NH algebra can be enlarged to the algebra $\widehat{NH}$ by the addition of the  generators $F_i$ 
that describe the constant acceleration transformations (see also [5]). 
These generators  satisfy the relation (\ref{oneb}) and
\begin{equation}
[F_i,\,F_j]\,=\,[F_i,\,K_j]\,=\,0,\quad [H,\,F_i]\,=\,2K_{i}.\label{four}
\end{equation}
By considering now the Jacobi identity for the generators $(H,P_i,F_j)$ one can show that in the 
acceleration-enlarged $\widehat{NH}_{\pm}$ algebra one must put $m=0$. However, one can show that in an arbitrary 
space-time dimension $D$ one can introduce one central charge $c$ as follows:
\begin{equation}
\label{five}
[K_i,\,F_j]\,=\,2c\delta_{ij}.
\end{equation}
The algebraic consistency then requires that
\begin{equation}
\label{six}
[P_i,\,K_j]\,=\,-\frac{c}{R^2}\delta_{ij}.
\end{equation}

In $D=(2+1)$ one can introduce, besides $c$, two additional central charges $\theta$ and $\theta'$.
The first one is already present in (\ref{threea}-b). % and additionally leads to the noncommutativity of the momenta:
%\begin{equation}
%\label{seven}
%[P_i,\,P_j]\,=\,-\frac{c}{R^2}\delta_{ij}.
%\end{equation}
The second central charge occurs only in the relation
\begin{equation}
\label{eight}
[F_i,\,F_j]\,=\,\theta'\epsilon_{ij}.
\end{equation}
However, it is easy to check that $\theta$ is a genuine central charge, determining the structure
of the enveloping algebra; the other two can be generated by the linear transformations
of the 
%vectorial $\widehat{NH}_{\pm}$
 generators $(P_i,K_i,F_i)$ in the enveloping algebra $U(\widehat{NH}_{\theta})$\footnote{By $\widehat{NH}_{\theta}$
we denote the accelaration-enlarged NH algebra with one central charge $\theta$.}
\begin{eqnarray} \nonumber
P_i\,&\rightarrow &\,\tilde P_i\,=\,\gamma P_i\,+\,\frac{c}{2\theta R^2\gamma} \epsilon_{ij}K_j,\\
\label{nine}
K_i\,&\rightarrow &\,\tilde K_i\,=\,\gamma K_i\,+\,\frac{c}{2\theta \gamma} \epsilon_{ij}P_j,\\
F_i\,&\rightarrow &\,\tilde F_i\,=\,\left(\gamma-\frac{\rho}{2R^2}\right)F_i\,+\,\frac{c}{\gamma \theta}\epsilon_{ij}K_j\,+\,\rho P_i,\nonumber
\end{eqnarray}
where
\begin{eqnarray}
\gamma^2\,&=&\,\frac{1}{2}\left[1+\left(1+(\frac{c}{\theta R})^2\right)^{\frac{1}{2}}\right],\nonumber \\
\rho\,&=&\,2\gamma R^2\left[1-\left(1+\frac{\theta'-\frac{c^2}{\theta \gamma}}{4\gamma^2\theta R^2}\right)^{\frac{1}{2}}\right].
\label{ten}
\end{eqnarray}

Note that as $R\rightarrow \infty$ we have
\begin{equation}
\label{eleven}
\gamma \rightarrow 1,\qquad \rho \rightarrow \frac{c^2}{4\theta^2}\,-\,\frac{\theta'}{4\theta}
\end{equation}
and so we see that our formulae (\ref{nine})-(\ref{ten}) generalise to the case $R<\infty$ the relevant expressions given in [1].

Finally, let us mention the two Casimirs of the accelaration enlarged Newton-Hooke algebra with central charges. They are
\bl
\begin{equation}
\label{twelvea}
C_H=H+\frac{c}{2\theta^2}\frac{1}{1+\frac{c^2}{\theta^2 R^2}}\left(\vec P^2-\frac{\vec{K}^2}{R^2}\right)
-\frac{1}{\theta(1+\frac{c^2}{\theta^2R^2})}\epsilon_{ij}K_iP_j
\end{equation}
and, puting $J_{ij}=\epsilon_{ij}J$ for $D=(2+1)$ 
\begin{equation}
\label{twelveb}
C_J\,=\,J\,-\,a\left(\vec F\vec P-\vec K^2-\frac{\vec F^2}{4R^2}\right)\,-\,\frac{c}{\theta}H\,-\,b\left(\vec P^2-\frac{\vec K^2}{R^2}\right),
\end{equation}
where
\begin{equation}
a\,=\,\frac{1}{2\theta+\frac{\theta'}{2R^2}},\qquad b\,=\,\frac{a\theta'}{4\theta}.\label{thirteen}\end{equation}
\el 

One can check that in the limit $R\rightarrow \infty$ these Casimirs reduce to the Casimirs of the acceleration-enlarged
Galilei algebra given in [1].

Note that in $D=(2+1)$ we can add to the above given $\widehat{NH}_+$ algebra with one central charge two further 
generators 
\begin{equation}
\label{fourteen}
J_{\pm}\,=\,\frac{1}{4\theta}\left(K_{\pm}^2-F_{\pm}P_{\pm}+\frac{F_{\pm}^2}{4R^2}\right),
\end{equation}
where $K_{\pm}=K_1\pm iK_2$ {\it etc}. If we now put $C_J=0$ and express $J$ from the formula (\ref{twelveb}), the triple
of generators $(J,J_+,J_-)$ provides a basis of a Lie algebra $O(2,1)$. If we now make the substitution
$R^2\rightarrow -R^2$ ({\it ie} ($\widehat{NH}_+\rightarrow \widehat{NH}_-$), as can be easily checked this algebra
becomes $O(3)$. In fact, an analogous result for the exotic  ((2+1) dimensional with two central charges) $NH$ algebras,
 in the case of $F_i=0$, was discussed in [11].

\section{$D=(2+1)$ Lagrangian models with acceleration-enlarged NH symmetries and with central charges}

When we put central charges to zero our algebra given by the relations (\ref{onea}-d) and (\ref{four}) can be realised by the following differential operators on the nonrelativistic $D=(2+1)$ space-time ($x_i,t)$ ($i=1,2)$
\footnote{Note that for finite $R$ suitable linear combinations of $P_i$
and $F_i$ generate standard translations.}:
\begin{eqnarray}
H\,=\,\frac{\partial}{\partial t},\qquad &&J\,=\, \epsilon_{ij}x_i\frac{\partial}{\partial x_j},\nonumber\\
P_i\,=\,\cosh\frac{t}{R}\,\frac{\partial}{\partial x_i},\quad && K_i\,=\,R\sinh\frac{t}{R}\,\frac{\partial}{\partial x_i},\label{fifteen}\\
F_i\,=\,&&\hspace{-0.5cm}2R^2(\cosh\frac{t}{R}-1)\frac{\partial}{\partial x_i}.\nonumber
\end{eqnarray}

Note that we have the following $\widehat{NH}$ transformations of the $D=(2+1)$ space-time ($\delta a_i$ - translations,
$\delta v_i$ - boosts, $\delta b_i$ - constant accelerations, $\delta a$ - time displacement and $\delta \alpha $ - $O(2)$ -
angle rotation)
\begin{eqnarray}
\delta t\, &=& \delta a,\nonumber \\
\delta x_i\,&=& \delta \alpha \epsilon_{ij}x_j\,+\,\cosh\frac{t}{R}\,\delta a_i \,+\label{sixteen}\\
&+& R\sinh\frac{t}{R}\,\delta v_i\,+\,2R^2\left(\cosh\frac{t}{R}\,-\,1\right)\,\delta b_i.\nonumber
\end{eqnarray}

Next we look for Lagrangians which are quasi-invariant under the transformations (\ref{sixteen}) and which, in the limit $R\rightarrow \infty$, reduce to the higher derivative Lagrangian given in [1], namely:
\begin{equation}
L_{R=\infty}(c,\theta,\theta')\,=\,-\frac{\theta}{2}\epsilon_{ij}\dot x_i\ddot x_j\,+\,\frac{c}{2}\ddot x_i^2\,-\,\frac{\theta'}{8}\epsilon_{ij}\ddot x_{i}\dot{\ddot{x}}_j.
\label{seventeen}
\end{equation}

Note that when $c=\theta'=0$ and adding, for $m\ne0$, the kinetic  term
$\frac{m}{2}\dot x_i^2$,  we get the Lagrangian considered in [13]. The extension of this Lagrangian to the NH case
was obtained in [8,11] via the substitution:
\begin{equation}
\label{eighteen}
\epsilon_{ij}\dot x_i\ddot x_j\,\rightarrow \epsilon_{ij}\left( \dot x_i\ddot x_j\,+\,\frac{1}{R^2}x_i\dot x_j\right)
\end{equation}

It is easy to check that this substitution can be also generalied to the other two terms in (\ref{seventeen}) as follows
\begin{eqnarray}
\ddot x_i^2\,&\rightarrow& \ddot x_i^2\,+\,\frac{1}{R^2}\dot x_i^2,\nonumber\\
\epsilon_{ij}\ddot x_i\dot{\ddot{x}}_j\,&\rightarrow&\epsilon_{ij}\left(\ddot x_i\dot{\ddot{x}}_j\,+\,\frac{1}{R^2}
\dot x_i\ddot x_j\right).\label{nineteen}
\end{eqnarray}

It is easy to check that the RHS terms in (\ref{eighteen}) and (\ref{nineteen}) are quasi-invariant under the transformations (\ref{sixteen}). If we perform the substitutions (\ref{nineteen}) together with
 $\theta\rightarrow \tilde \theta=\theta + \frac{\theta'}{4R^2}$ we obtain
\begin{eqnarray}
L_{NH}\,&=&\,-\frac{\tilde\theta}{2}\,\epsilon_{ij}\left(\,\dot x_i\ddot x_j\,+\,\frac{1}{R^2} x_i\dot x_j\right)\,+
\frac{c}{2}\left(\ddot x_i^2\,+\,\frac{1}{R^2}\dot x_i^2\right)-
\nonumber \\
&-&\,\frac{\theta'}{8}\epsilon_{ij}\,\left(\ddot
 x_i\dot{\ddot{x}}_j\,+\,\frac{1}{R^2}\dot x_i\ddot x_j\right).\label{twenty}
\end{eqnarray}

% So, we have as our Lagrangian
%\begin{eqnarray}
%L_{NH}\,&=&\,\frac{\theta}{2R^2}\,\epsilon_{ij}\,x_i\dot x_j\,-\,\frac{\tilde \theta}{2}\epsilon_{ij}\,\dot x_i\ddot x_j\,+,
%\nonumber \\
%&+&\frac{c}{2}\left(\ddot x_i^2\,+\,\frac{1}{R^2}\dot x_i^2\right)\,-\,\frac{\theta'}{8}\epsilon_{ij}\,\ddot
 %x_i\dot{\ddot{x}}_j,\label{twenty}
%\end{eqnarray}

Note that the substitution $\theta\rightarrow \tilde \theta$ is necessary to get agreement with the definition of the 
central charge $\theta$ in (\ref{threea}).

When, below,  we rewrite our Lagrangian (\ref{twenty}) in the first order Hamiltonian formalism we need to introduce five pairs of
variables $(x_i,p_i,y_i,q_i,u_i)$.
\section{First order Hamiltonian formalism and the extended phase space}

First we define two new 2-vector variables by
\begin{equation}
 \label{etwenty}
y_i\,=\,\dot x_i, \qquad u_i\,=\,\dot y_i.
\end{equation}
Then we obtain the following first order Lagrangian \footnote{ 
The first order quasi-invariant actions can be also derived by a
geometric technique which involves the method of nonlinear realisations of the
$\widehat{NH}_{+}$ group with central extensions and the inverse Higgs mechanism
(for the $NH_{+}$ group see [11]; see also [14]). It would be interesting to
describe the most general class of first order actions that can be generated this way.}
\begin{eqnarray}
L_{NH}\,&&=\,p_i(\dot x_i-y_i)\,+\,q_i(\dot y_i-u_i)\,-\,\frac{\tilde\theta}{2}\epsilon_{ij}y_i\dot y_j\,-\,\frac{\theta'}{8}
\epsilon_{ij}u_i\dot u_j\nonumber\\
&& +\frac{c}{2}\left(u_i^2+\frac{y_i^2}{R^2}\right)\,-\,\frac{\tilde\theta}{2R^2}\epsilon_{ij}x_iy_j\,-\,\frac{\theta'}{8R^2}\epsilon_{ij}y_iu_j
\label{etwentyone}
\end{eqnarray}
which, after the substitution (\ref{etwenty}), is  equivalent to the higher order action (\ref{twenty}).
Using the Faddeev-Jackiw prescription [15,16] we obtain the following non-vanishing Poisson brackets:
\begin{eqnarray}
 \{x_i,\,p_j\}\,=\,\delta_{ij},&\quad& \{y_i,\,q_j\}\,=\,\delta_{ij}\nonumber\\
\{q_i,\,q_j\}\,=\,-\tilde\theta \epsilon_{ij},&\quad&\{u_i,\,u_j\}\,=\,\frac{4}{\theta'}\epsilon_{ij}.
\label{etwentytwo}
\end{eqnarray}

The Hamiltonian, which follows from (\ref{etwentyone}) has the form
\begin{equation}
 H=p_iy_i+q_iu_i-\frac{c}{2}\left(u_i^2+\frac{1}{R^2}y_i^2\right)+\frac{\tilde\theta}{2R^2}\epsilon_{ij}x_iy_j+\frac{\theta'}{8R^2}\epsilon_{ij}y_iu_j.
\label{etwentythree}
\end{equation}
The equations of motion, which can be obtained from (\ref{etwentyone}) as the Euler-Lagrange equations or as the Hamilton
equations following from (\ref{etwentytwo}-\ref{etwentythree}), give, besides the equations (\ref{etwenty}) also
\bl
\begin{eqnarray}
\dot u_i\,&=&\,\frac{1}{2R^2}y_i\,+\,\frac{4}{\theta'}\epsilon_{ij}(q_j-cy_j)\\
\dot q_i\,&=&\,-\left(\tilde\theta+\frac{\theta'}{8R^2}\right)\epsilon_{ij}u_j\,+\,\frac{\tilde \theta}{2R^2}\epsilon_{ij}x_j\,+\,\frac{c}{R^2}y_i\,-\,p_i\label{fourb}\\
\dot p_i\,&=&\,-\frac{\tilde \theta}{2R^2}\,\epsilon_{ij}y_j.\label{fourc}
 \end{eqnarray}
\el
Using the first two equations as defining the variables $q_j$, $p_i$ we obtain the following set of the $\widehat{NH}$ transformation laws for the extended phase space coordinates: $(x_i,p_i,y_i,q_i,u_i)$:

\begin{itemize}
 \item $\widehat{NH}_+$ translations (parameters $a_i$)
\begin{eqnarray}
 \delta x_i\,&=&\, a_i \cosh\frac{t}{R}, \qquad\quad \delta y_i\,=\,\frac{a_i}{R}\sinh\frac{t}{R},\nonumber \\
\delta u_i\,&=&\,\frac{a_i}{R^2}\cosh\frac{t}{R},\,\qquad\quad \delta p_i\,=\,-\epsilon_{ij}a_j\frac{\tilde \theta}{2R^2}\cosh\frac{t}{R},\label{etwentyfive}\\
\delta q_i\,&=&\, a_i\frac{c}{R^2}\cosh\frac{t}{R}\,-\,\epsilon_{ij}a_j\frac{\theta'}{8R^3}\sinh\frac{t}{R}.\nonumber
\end{eqnarray}
\item $\widehat{NH}_+$ boosts (parameters $v_i$)
\begin{eqnarray}
 \delta x_i\,&=&\, v_i R \sinh\frac{t}{R}, \qquad\quad \delta y_i\,=\,v_i\cosh\frac{t}{R},\nonumber \\
\delta u_i\,&=&\,\frac{v_i}{R}\sinh\frac{t}{R},\,\qquad\quad \delta p_i\,=\,-\epsilon_{ij}v_j\frac{\tilde \theta}{2R}\sinh\frac{t}{R},\label{etwentysix}\\
\delta q_i\,&=&\, \frac{v_ic}{R}\sinh\frac{t}{R}\,-\,\epsilon_{ij}v_j\frac{\theta'}{8R^2}\cosh\frac{t}{R}.\nonumber
\end{eqnarray}
\item $\widehat{NH}_+$ accelerations (parameters $b_i$)
\begin{eqnarray}
 \delta x_i\,&=&\, 2b_i R ^2\left(\cosh\frac{t}{R}-1\right),\qquad \quad \delta y_i\,=\,2b_iR\sinh\frac{t}{R},\nonumber \\
\delta u_i\,&=&\,2b_i\cosh\frac{t}{R},\,\qquad\quad \delta p_i\,=\,-\epsilon_{ij}b_j\tilde \theta\left(\cosh\frac{t}{R}+1\right),\label{etwentyseven}\\
\delta q_i\,&=&\, 2b_ic\cosh\frac{t}{R}\,-\,\epsilon_{ij}b_j\frac{\theta'}{4R^2}\sinh\frac{t}{R}.\nonumber
\end{eqnarray}
\end{itemize}

We leave it as an exercise for the interested reader to determine the Noether charges generating (\ref{etwentyfive}-\ref{etwentyseven}).
They will satisfy the acceleration-enlarged $\widehat{NH}$-algebra with central charges introduced in Section 2.

\section{``External'' and ``internal'' dynamics sectors with noncommutative structure}

%It is easy to check that the following linear combinations of the phase-space variables are invariant under
% the transformations %(\ref{etwentyfive}-\ref{etwentyseven}):
Next we look for internal variables which are invariant under the transformations (\ref{etwentyfive}-\ref{etwentyseven})
and so could be used to parametrise the internal sector of the phase space with the time evalution being an internal
automorphism. We note that for these internal sector variables we can take 
\begin{eqnarray}
 U_i\,&=&\,u_i\,-\,\frac{1}{\tilde \theta}\epsilon_{ij}\left(p_j\,+\,\frac{\tilde \theta}{2R^2}\epsilon_{jk}x_k\right)\,-\,\frac{x_i}{R^2}\nonumber \\
Q_i\,&=&\,q_i\,-\,cu_i\,+\,\frac{\theta'}{8R^2}\epsilon_{ij}y_j\label{etwentyeight}
\end{eqnarray}
as they are invariant under the transformations (\ref{etwentyfive}-\ref{etwentyseven}).

From eq. (\ref{etwenty}) and (23a-c) we get 
\begin{eqnarray}
 \dot U_i\,&=&\, \frac{4}{\theta'}\epsilon_{ij}Q_j,\nonumber\\
\dot Q_i\,&=&\,-\frac{4c}{\theta'}\epsilon_{ij}Q_j\,-\,\tilde\theta \epsilon_{ij}U_j.
\label{etwentynine}
\end{eqnarray}

Hence we see that the dynamics of the variables $U_i$ and $Q_i$ is closed under the time evolution, and as such, it describes
the ``internal'' dynamics in our model. Let us note that
\begin{itemize}
 \item The set of equations (\ref{etwentynine}) leads to the fourth order equations for the invididual components of 
the fields $U_i$ and $Q_i$.
\item The description of the dynamics of the internal sector ($U_i,Q_i)$ does not depend on $R$, {\it i.e.} it is the same in the Galilean 
limit ($R\rightarrow \infty$) considered in [1] after the substitution $\theta\rightarrow \tilde\theta$ has been made.
\end{itemize}

To describe the ``external'' sector of our model we have to introduce three sets of variables ($X_i, P_i,Y_i$)  which have
vanishing Poisson brackets with $U_i$ and $Q_i$. Then the two sectors will be dynamically independent.

Note that from (\ref{etwentyeight}) and (\ref{etwentytwo})  we have
\begin{eqnarray}
 \{U_i,\,U_j\}\,=\,4\frac{\theta}{\theta'\tilde\theta}\epsilon_{ij},&&\qquad \{Q_i,\,U_j\}\,=\,-\frac{4c}{\theta'}\epsilon_{ij}\nonumber\\
\{Q_i,\,Q_j\}&&=\,\epsilon_{ij}\left(-\tilde\theta+\frac{4c^2}{\theta'}+\frac{\theta'}{4R^2}\right).
\label{ethirty}
\end{eqnarray}
Then, if we define 
\begin{equation}
 X_i\,=\,x_i\,+\,\alpha Q_i\,+\,\beta U_i\label{ethirtyone}
\end{equation}
we find that
\begin{equation}
 \{X_i,Q_j\}=\{X_i,U_j\}\,=\,0
\end{equation}
are satified if
\begin{equation}
 \alpha=\left(\frac{\theta^2}{c}+\frac{c}{R^2}\right)^{-1},\qquad \beta=\left(c-\frac{\theta\theta'}{4c}\right)\alpha.
\end{equation}

For $Y_i$ we try the ansatz $Y_i=\dot X_i$. Then using the Jacobi identity 
 and the field equations (\ref{etwentynine}) we see that\footnote{In a similar way we can show that $\{Y_i,Q_j\}=0$}
\begin{eqnarray}
 \{Y_i,\,U_j\}\,=&&\{\{X_i,\,H\},U_j\}\,=\,-\{X_i,\,\{Q_j,\,H\}\}\nonumber\\
=&&-\frac{4}{\theta'}\epsilon_{ij}\,\{X_i,Q_j\}\,=\,0.\label{ethirtythree}
\end{eqnarray}
Thus we see that we can put $Y_i=\dot X_i$, or we can use the field equations (29) and note that
\begin{equation}
 Y_i\,=\,y_i\,-\,\alpha\left(\tilde \theta \epsilon_{ij}U_j\,+\,\frac{\theta}{c}\epsilon_{ij}Q_j\right).
\label{ethirtyfour}
\end{equation}

To find the third set of variables of the external variables we calculate the time derivative of $Y_i$.
We find 
\begin{equation} 
 \dot Y_i\,=\,\frac{X_i}{R^2}\,+\,\frac{1}{\tilde \theta}\epsilon_{ij}P_j,
\end{equation}
where $P_i$ is given by
\begin{equation}
 P_i\,=\,p_i\,+\,\frac{\tilde \theta}{2R^2}\epsilon_{ij}x_j.
\end{equation}

Note that
\begin{itemize}
 \item The variables $P_i$ are invariant under the $\widehat{NH}_+$ translations and $\widehat{NH}_+$ boosts.
\item The ``external'' 2-momenta $P_i$ are constant ``on-shell'', {\it i.e.} from (36)
 (\ref{etwenty}), and (23c) we see that
\begin{equation}
\dot P_i\,=\,0.
\end{equation}
\end{itemize}

%The dynamics of the external sector is described by the equation of the inhomogeneous hyperbolic oscillator:
%\begin{equation}
% \ddot X_i\,-\,\frac{1}{R^2}X_i\,=\,\frac{1}{\tilde \theta}\epsilon_{ij}P_j,
%\end{equation}
%where the variable on the rhs of the equation above is constant. 
The equal time non-commutative 
structure of the ``external'' sector variables $X_i, P_i, Y_i=\dot X_i$ is determined by the following set of Poisson brackets:
\begin{eqnarray}
 \{X_i,\,P_j\}\,=\,\delta_{ij},\qquad&&\{P_i,\,P_j\}\,=\,\frac{\tilde\theta}{R^2}\epsilon_{ij}\nonumber\\
\{X_i,\,X_j\}\,=\,\frac{\beta}{\tilde\theta}\epsilon_{ij},\qquad&&\{X_i,\,Y_j\}\,=\,-\alpha\frac{\theta}{c}\epsilon_{ij}\label{eth}\\
\{Y_i,\,Y_j\}\,=\,\alpha\frac{\theta}{c}\epsilon_{ij},\qquad&&\{Y_i,\,P_j\}\,=\,0\nonumber.
\end{eqnarray}
Therefore, by means of the substitution $
 X_i\,\rightarrow \, X_i'\,=\,X_i\,+\,\frac{R^2}{\tilde\theta}\epsilon_{ik}P_k
$ we obtain the equations of the hyperbolic oscillator
\begin{equation}
 \ddot X_i'\,=\,\frac{1}{R^2}X_i'
\end{equation}
and a decoupling of $P_i$ from ($X_i',Y_i)$ in the Poisson Brackets (38).

We see that our present model differs in many respects from the original
higher-derivative model [13,17]:
%in comparison with our model given in [12,16], which can be obtained from (19) by putting $\theta'=c=0$, we have the following
%modifications:
\begin{itemize}
 \item The external sector, described in [13] by a pair ($X_i,P_i)$ of non-commutative $X_i$ and commutative $P_i$ variables, is replaced now by a triplet of variables
($X_i,P_i,Y_i)$ with both $X_i$ and $P_i$ being noncommutative (see formula (38)).
\item The internal sector described in [13] by a single 2-vector is spanned in the present model by
a pair of 2-vectors ($Q_i,U_i)$ with the dynamics described by the equations with the fourth order time derivatives (see {\it e.g.} [18,19]).
However, after a complex point transformation $(Q_i,U_i)\rightarrow (z_i,\pi_i)$ the internal part $L_{int}$ of the first order Lagrangian (20) can be put into the standard form
with a canonical symplectic structure $(L_{int}=\dot z_i\pi-H_{int}(z_i,\pi_i))$ (cp [20]).

\end{itemize}

\section{Final Remarks}

This paper has had three aims:
\begin{itemize}
 \item To show that the acceleration-enlarged Galilean symmetries can be generalised to the acceleration-enlarged Newton-Hooke
symmetries describing very special nonrelativisitic de-Sitter geometries. In particular we have shown that the acceleration-enlarged $\widehat{NH}$
algebra in ($D=2+1$) dimensions has three central charges.
\item It has been known for some time that mass, as a central charge of the Galilean symmetry, provides a unique parameters
which controls the free particle motion. Now we see that in the $D=(2+1)$ dimensions the $\widehat{NH}$ invariant action depends on three 
parameters which are determined by three central charges of the accelaration enlarged $\widehat{NH}$ algebra.
\item It was shown in [13] that in $D=(2+1)$ dimensions the exotic central charge of the Galilean algebra generates the noncommutativity of the $D=2$ nonrelativistic
space coordinates. In this paper we have shown that for a system with $\widehat{NH}$ symmetries (dS radius $R$ finite)
one can generate a more complicated nonrelativistic $D=2$ phase space with both coordinates and momenta being noncommutative..
We have found also that the complete noncommutative phase space describing the external sector of our model, besides $(X_i,P_i)$, possesses also a third non-commutative 2-vector
\end{itemize}

As is well known the quantization of models with higher time derivatives leads to the canonical derivation of the deformed quantum phase space structures. 
It still remains  to be clarified how the well known difficulty with the quantization of mechanical systems with higher order equations of motion, leading to the appearance of ghosts,
can be remedied if we formulate the model in the framework of suitably extended and deformed quantum mechanics.

{\bf Acknowledgments}: We would like to thank Jose de Azcarraga, Joaquim Gomis  and Mikhail Plyushchay for  valuable remarks.

\end{document}